\documentclass[referee]{aa}
\usepackage[varg]{txfonts}
\usepackage{graphicx}
\usepackage{natbib}
\bibpunct{(}{)}{;}{a}{}{,} % to follow the A&A style
\newcommand{\al}{$^{26}$Al }

\begin{document}

\title{Solar energetic particles and galactic cosmic rays over millions of years as inferred from data on cosmogenic $^{26}$Al in lunar samples}

\author{S. Poluianov\inst{1,2}
\and G. A. Kovaltsov\inst{3}
\and I.G. Usoskin\inst{1,2}
}

\institute{Space Climate Research Unit, University of Oulu, Finland
\and Sodankyl\"a Geophysical Observatory, University of Oulu, Finland
\and Ioffe Physical-Technical Institute, 194021 St. Petersburg, Russia
}

\date{}

\abstract {}
%Aims
{Lunar soil and rocks are not protected by a magnetic field or an atmosphere and are continuously irradiated by
 energetic particles  that can produce cosmogenic radioisotopes directly inside rocks at different depths
 depending on the particle's energy.
This allows  the mean fluxes of solar and galactic cosmic rays to be assessed on the very long timescales of millions of years.
}
%Methods
{Here we show that lunar rocks can serve as a very good particle integral spectrometer in the energy range 20\,--\,80 MeV.
We have developed a new method based on precise modeling, that is applied to measurements of \al (half-life $\approx 0.7$ megayears)
 in lunar samples from the \textit{Apollo} mission, and present the first direct reconstruction (i.e., without any {a priori}
 assumptions) of the mean energy spectrum of solar and galactic energetic particles over a million of years.}
%Results
{We show that the reconstructed spectrum of solar energetic particles is totally consistent with that over the last decades,
 despite the very different levels of  solar modulation of galactic cosmic rays ($\phi=496\pm40$ MV over a million years
  versus  $\phi=660\pm 20$ MV for the modern epoch).
We also estimated the occurrence probability of extreme solar events and argue that no events
 with the $F(>30$ MeV) fluence exceeding $5\cdot 10^{10}$ and $10^{11}$ cm$^{-2}$ are expected
 on  timescales of a thousand and million years, respectively.}
%Conclusions
{We conclude that the mean flux of solar energetic particles hardly depends on the level of solar activity,
 in contrast to the solar modulation of galactic cosmic rays.
This puts new observational constraints on solar physics and becomes important for
 assessing radiation hazards for the planned space missions.
}

\keywords{Sun:particle emission - Sun:activity - Moon}
\titlerunning{SEP on million year timescales}
\maketitle

%%.-.-.-.-.-.-.-.-.-.-.-.-.-.-.-.-.-.-.-.-.-.-.-.-.-.-.-.-.-.
%.........................
\section{Introduction}

Solar energetic particles (SEPs), which form an essential component of the radiation environment near Earth
 \citep{vainio09,schwadron17},
 appear as sporadic fluxes of energetic protons and a small fraction of heavier particles associated with powerful
 solar flares and/or coronal mass ejections.
Knowledge of the mean SEP flux in the energy range of several tens of MeV, as well as an assessment of the strength of occurrence
 probability of extreme SEP events are crucially important for the modern space-based technological society \citep[e.g.,][]{NCRP06}.
One of the most important parameters of the near-Earth radiation environment is the integral flux of energetic particles
 with energy above 30 MeV, $F(>$30 MeV) \citep[e.g.,][]{shea90,feynman93}.
Direct space-borne measurements of SEPs over the last decades suggest that the average SEP flux is dominated by rare major
 events \citep{bazilevskaya14} and varies between individual solar cycles by an order of magnitude \citep{reedy12}.
However, the direct data were obtained during the period of unusually high solar activity known as the modern
 grand maximum \citep[MGM, see][]{solanki_nat_04}.
Therefore, it is unclear whether the modern data are representative for longer timescales.

Extreme SEP events in the past can be studied by means of cosmogenic isotopes
 (primarily $^{14}$C and $^{10}$Be) in terrestrial natural archives
 \citep[e.g.,][]{miyake12, guettler15, usoskin_LR_17} on a timescale of up to 10000 years; however,   because of
  atmospheric and magnetospheric shielding, they cannot provide
 information about solar particles with the energy of tens of MeV \citep{kovaltsov_F200} that is most important
 for technological impacts.
Ideally, this ought to be studied using data from the  space outside the Earth's atmosphere and
 magnetosphere, such as meteorites \citep[e.g.,][]{mancuso18} or lunar samples.
The lunar surface is not shielded from incoming radiation and keeps the information about its flux in the past.
Cosmogenic isotopes are produced in situ in lunar rocks, allowing us to reconstruct the mean energy spectrum of
 energetic particles, although without temporal resolution.
This idea was explored earlier \citep{reedy72} when the SEP flux was estimated using
 measurements of lunar rocks brought to Earth by the \textit{Apollo} missions.
However, the earlier efforts \citep[e.g.,][]{rao94,fink98, jull98, nishiizumi09} were model-dependent.
In fact, those works did not provide true reconstructions of the spectrum, but only estimated the parameters of an explicitly prescribed functional shape
 (exponential over rigidity; see Eq.~\ref{eq:exp}).
Here we demonstrate that a lunar rock with  cosmogenic \al\ produced in situ  can serve as an integral particle spectrometer
 able to reconstruct the SEP energy spectrum over long-term scales directly from measurements without any
 {a priori} assumptions on the spectral shape.

\section{Data and methods}
%.........................
\subsection{Parameters of the used lunar samples}
\label{Sec:data}

Here we use data of \al (half-life 717000 years) activity in two lunar samples, 64455 and 74275,
 brought by Apollo missions 16 and 17, and subsequently measured in different laboratories.

%\subsection{Sample 64455}
Sample 64455 is an egg-shaped object of about 5 cm long and 3 cm across collected directly from the lunar surface.
Full information on its physical and chemical parameters is available
 elsewhere\footnote{\textsf{http://curator.jsc.nasa.gov/lunar/lsc/64455.pdf}} \citep{meyer11}.
The content of \al was measured by \citet{nishiizumi09}.
The sample was modeled  using hemispherically concentric shells ($R=7$ g/cm$^2$) lying upon
 a flat lunar-soil surface.
The irradiation age and erosion rate were estimated as 2 Myr and 0\,--\,0.5 mm/Myr, respectively \citep{meyer11}.

%\subsection{Sample 74265}
Sample 74275 is a flat knob-shaped object, 17 cm long,   12 cm across, and 4 cm thick, collected from the lunar surface.
Full information is available elsewhere\footnote{\textsf{http://curator.jsc.nasa.gov/lunar/lsc/74275.pdf}} \citep{meyer11}.
It was measured by \citet{fink98} and modeled using ``knob'' geometry ($R=78$ g/cm$^2$) and columnar-averaged isotope activity,
 as lying upon a flat lunar-soil surface.
The irradiation age and erosion rate were estimated as 2.8 Myr and 1\,--\,2 mm/Myr, respectively \citep{meyer11}.

We also use \al measurements \citep{rancitelli75,nishiizumi84} in the 242-cm long \textit{Apollo 15} deep-drill core,
 which was modeled using the slab geometry and chemical composition according to \citet{gold77}.
More detailed information is available\footnote{\textsf{https://curator.jsc.nasa.gov/lunar/lsc/A15drill.pdf}} at \citep{meyer11}.

%.........................
\subsection{Model of $^{26}$Al production}
\label{s:Al26_prod}

Production of an isotope by cosmic rays inside matter is numerically modeled using the concept of the yield function.
The isotope's production rate $Q(h)$ at depth $h$ in a sample is related to the flux of primary energetic particles,
 SEPs or galactic cosmic rays (GCRs), as
\begin{equation}
Q(h)=\sum_i \int Y_i(E,h)\cdot J_i(E)\cdot dE,
\label{eq:Q}
\end{equation}
where index $i$ denotes the type of primary energetic particles, i.e., protons, $\alpha$-particles,  or heavier species;
 $Y_i(E,h)$ is the isotope yield function; and $J_i(E)$ is
 the differential intensity of particles with energy $E$.
The intensity $J$ is typically used to quantify GCRs in units of [cm$^2$\,s\,sr\,MeV]$^{-1}$, while for SEPs the omnidirectional
 flux $F$ [cm$^2$\,s\,MeV]$^{-1}$ is often used, which is related to the intensity, in the isotropic case, as
 $F(E) = 4\pi\cdot J(E)$ \citep[see, e.g., chapter 1.6 of][]{grieder01}.

\begin{figure}
\begin{center}
\includegraphics[width=1\columnwidth]{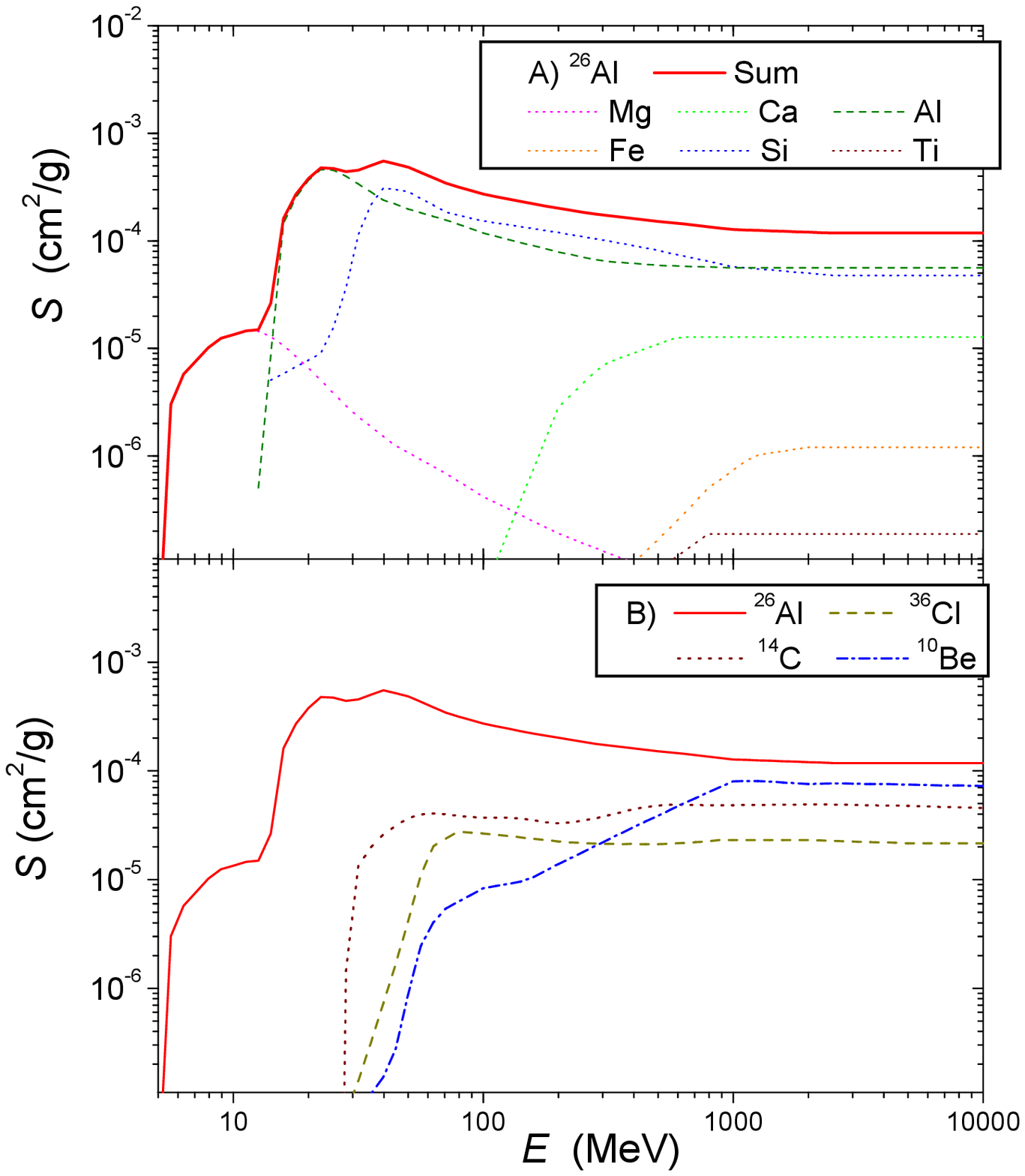}
\end{center}
\caption{
Efficiency of production (see Eq.~\ref{eq:prod_eff}) of cosmogenic isotopes by
 primary and secondary protons of energy $E$ in the lunar sample 64555, as computed here.
Panel A shows the production of \al: contributions from different target nuclei are denoted by dashed lines,
 while the thick red curve is the total sum.
Panel B depicts the summary productions of different long-lived cosmogenic isotopes, as denoted in the legend.
The red curve is identical to that in panel A.
}
\label{f:prod_eff}
\end{figure}

The yield function is calculated as
\begin{equation}
Y_i(E,h) = \sum_k \int S_k(E')\cdot N_{k,i}(E,E',h)\cdot v_k(E')\cdot dE',
\end{equation}
where $k$ is the index of primary or secondary particles (protons, neutrons, $\alpha$-particles, pions),
 $S_k(E')$ is the efficiency of the isotope production by a particle of type $k$ with energy $E'$,
 $N_{k,i}(E,E',h)$ is the concentration at depth $h$ of primary and secondary particles of type $k$ with energy $E'$
 corresponding to the primary particle of type $i$ with energy $E$, and $v_k(E')$ is the particle's velocity.
$N_{k,i}(E,E',h)$ is a result of numerical simulations of the particle transport in
 lunar matter and is calculated per unit intensity ($J_i=$1~cm$^{-2}\,$s$^{-1}$\,sr$^{-1}$) of monoenergetic primary particles.
The efficiency of isotope production by particles of type $k$ is
\begin{equation}
S_k(E') = \sum_j C_j\cdot \sigma_{k,j} (E'),
\label{eq:prod_eff}
\end{equation}
where $j$ indicates the type of target nuclei in a lunar sample,
 $C_j$ is their content in one gram of matter, and $\sigma_{j,k}(E')$ is the cross section  of
 the corresponding nuclear reactions.
The efficiency $S$ depends on the chemical composition of a sample and needs to be calculated for each case individually.
The cross sections of \al production by protons were adopted from \citet{nishiizumi09} and \citet{reedy07},
 while those for $\alpha$-particles were taken according to \citet{tatischeff06}.
The cross sections for \al production by neutrons on Al and Si were taken from \citet{reedy13} and assumed to be identical
 to those for the proton reactions for other nuclei (Ca, Ti, Fe).
We also considered the \al production by charged pions, whose contribution was typically neglected earlier but
 is shown to be essential in dense matter \citep{Li_lunar_17}.
The corresponding cross sections were extracted from the Geant4 model \citep{Geant4-1}.
An example of the efficiency of \al production for proton reactions in a lunar rock is shown in Figure \ref{f:prod_eff}.
We can see that three elements (Mg, Al, and Si) dominate the production in different energy ranges.
The isotope \al enables the most accurate reconstruction of the SEP spectrum due to the low energy threshold and
 high efficiency of its production by low-energy particles (Figure \ref{f:prod_eff}A), in contrast to other isotopes
  $^{10}$Be, $^{14}$C, and $^{36}$Cl, also measured in lunar rocks but having higher energy thresholds,
  lower efficiency (Figure \ref{f:prod_eff}B), and accordingly lower sensitivity to SEPs versus GCRs,
  even in the upper layers \citep{fink98, jull98, nishiizumi09}.

Specific yield functions were calculated individually for each particular case considered here (two lunar rocks
 and the deep core; see Section~\ref{Sec:data}).
Primary particles were assumed to arrive isotropically from the upper hemisphere.
Their composition was considered to be only protons for SEPs, and protons and heavier species for GCRs.
An example of the yield function for \al at different depths is shown in Figure \ref{f:yf}.

Computations of the isotope production in the deep-drill core require simulations of the nucleonic cascade
 initiated by energetic particles deeper in matter.
This was made by full Monte Carlo simulations, using the Geant4 toolkit \citep[version 4.10][]{Geant4-1}
 for nuclear interactions of primary and secondary particles in the core and applying the slab geometry.
The chemical composition of the  Apollo 15 deep-drill core was taken according to \citep{gold77}.

It was proposed earlier \citep{reedy72} that SEP related production in lunar rocks can be modeled analytically
 since SEP energy is insufficient to initiate a nucleonic cascade.
We have checked this assumption \citep{poluianov_ICRC_2015} and found that
 it works reasonably well for the upper layers shallower than 5\,--\,7 g/cm$^2$.
Accordingly, we simulated \al production by SEP protons in the shallow layers by direct integration of
 the analytical formulas for the exact geometry in each case.
Samples were considered to be  lying on typical lunar soil with a flat surface and the chemical composition of
 the Apollo 15 deep-drill core.

\subsection{Lunar rock as an integral spectrometer}
\label{Sec:spec}
\begin{figure*}
\begin{center}
\includegraphics[width=1\textwidth]{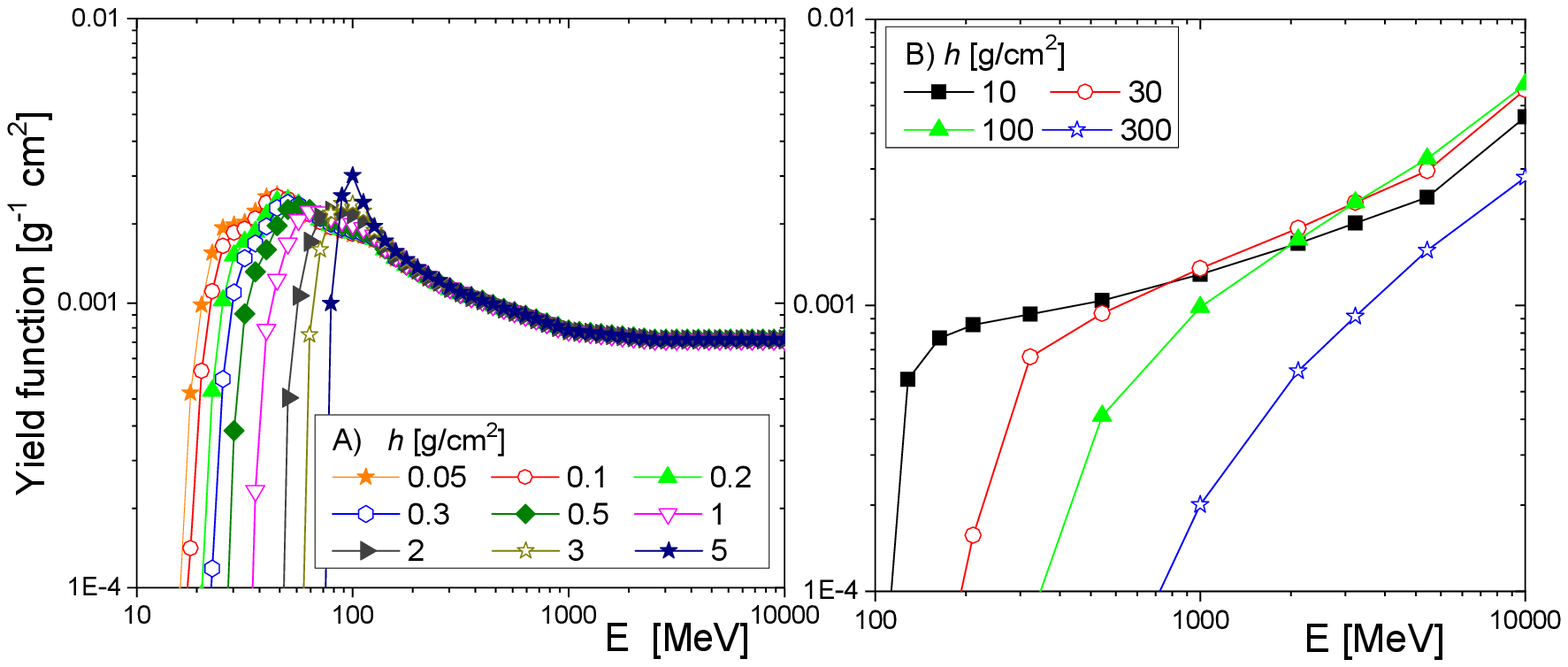}
\end{center}
\caption{
Examples of the yield function of \al production in the lunar rock for different depths as denoted in
 the legend in units of g/cm$^2$.
Panel A corresponds to shallow layers of sample 64455 in idealized conditions
 (infinite exposition age, no erosion).
Panel B corresponds to the Apollo 15 long drill core.
These yield functions are shown here for illustration, but for each case they were computed
 individually considering realistic exposition age, erosion rate, geometry, chemical composition, and
 measurement details (see Section~\ref{Sec:data}).
}
\label{f:yf}
\end{figure*}

The yield function for \al produced in lunar rocks by protons at shallow depths ($<$7 g/cm$^2$, Figure~\ref{f:yf}A) is step-like with the
 sharp energy threshold growing with depth, where \al is produced directly by impinging energetic particles as a result of
 spallation of target nuclei Mg, Si, and Al (Figure \ref{f:prod_eff}A) without development of a nuclear cascade.
Spallation reactions define the primary particle's energy threshold, which increases with depth
 because of ionization losses.
This yield-function shape is close to that of an ideal integral particle spectrometer, which is a detector whose response
 is directly proportional to the integral flux of
 primary particles with energy above a threshold $E^*$, with the yield-function being step-like, i.e., zero for
 energy below $E^*$, and a constant value for $E\geq E^*$.

\subsubsection{SEP spectral shape}
Here we show that production of \al in the upper layer of a lunar rock is indeed a good integral spectrometer for
 protons of SEP energies.
This allows the SEP spectrum to be reconstructed directly from the measured data, without
 any {a priori} assumption on its exact spectral shape.
To check this and to assess the related uncertainties, we considered two spectral shapes often used in earlier SEP studies:
 an exponent over rigidity (hereafter  EXP) \citep{freier63}
\begin{equation}
F(R)=F_0 \exp\left(-\,\frac{R}{R_0}\right)
\label{eq:exp}
,\end{equation}
and a power law (POW) over energy \citep{vanhollebeke75}
\begin{equation}
F(E)=F_0 E^{-\gamma}.
\label{eq:pow}
\end{equation}

Overall, POW tends to provide a spectrum that is  too hard with an excess of high- and low-energy particles,
 while EXP yields a softer (at the high-energy tail) spectrum with an excess of mid-energy particles.
These two models bound more realistic cases, such as the modified power law \citep{cramp97}, i.e.,  a power law
 with a gradually increasing spectral index; the Ellison--Ramaty spectrum \citep{ellison85},
 a power law with an exponential roll-off at higher energies;  Band-function \citep{band93,raukunen18},
 a double power law, hard at low energies and soft at higher energies with a smooth junction;
 or Weibull  representations \citep{pallocchia17} of the SEP energy spectrum.
An agreement between the results based on the POW and EXP models would imply that the method is model-independent,
 while the difference between them can serve as a conservative measure of the model uncertainty.

\subsubsection{Effective energy of isotope production}
\begin{figure*}
\begin{center}
\includegraphics[width=1\textwidth]{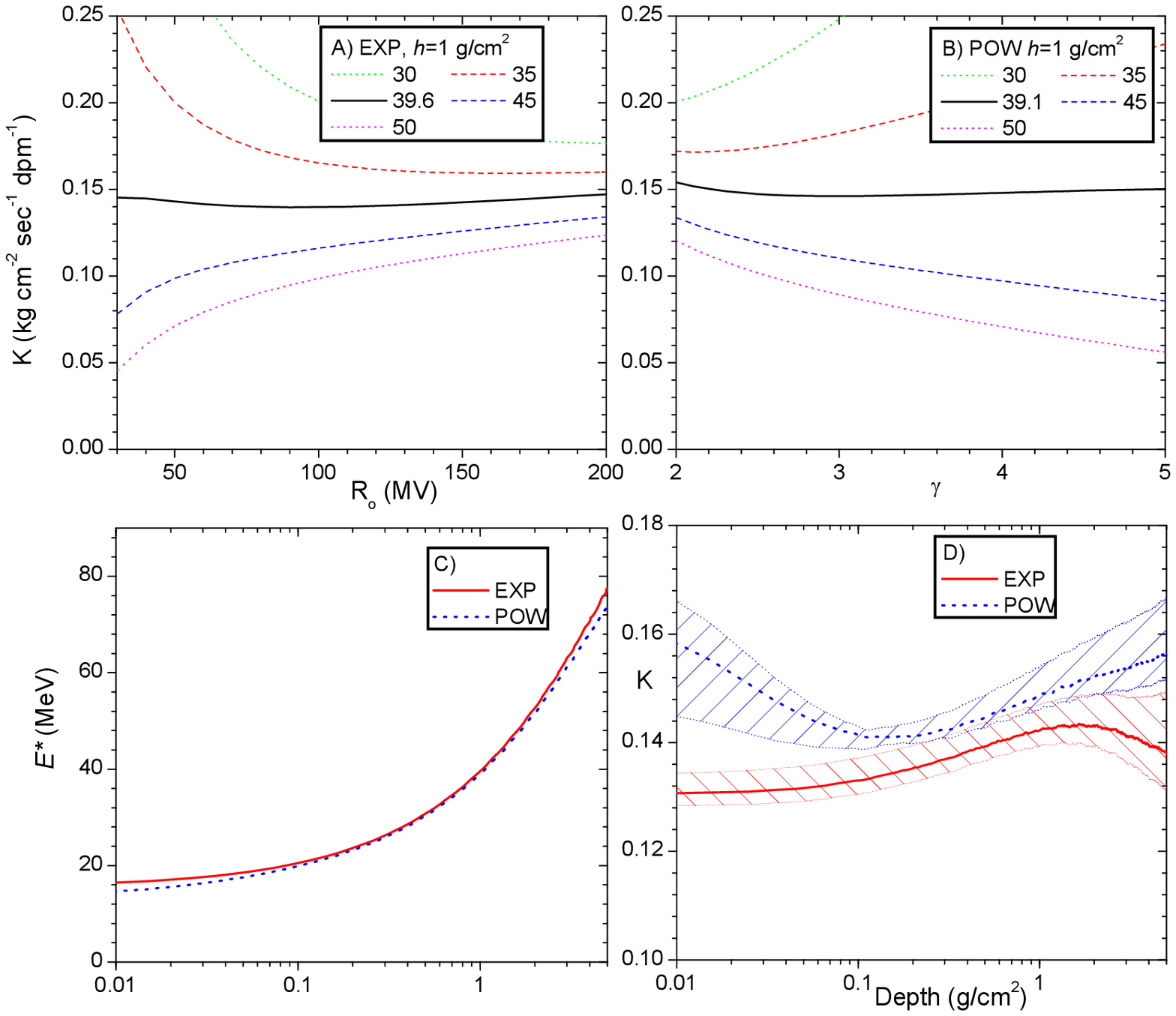}
\end{center}
\caption{
A) Dependence of the scaling factor $K$ on the characteristic rigidity $R_0$ of the EXP model
 for the depth of 1 g/cm$^2$ for different values of the effective energy $E^*$, as denoted in the legend (in MeV).
B) Same as A, but for the POW model with the index $\gamma$.
C) Effective energy $E^*$ as a function of  depth in lunar sample 64455 for the EXP and POW models.
D)  Same as C, but for the best-fit scaling factor $K$;  the hatched areas represent
 the full-range $\Delta K$-values.
}
\label{f:K_factor}
\end{figure*}

Here we define the effective energy $E^*$ of the isotope production \citep[see][]{kovaltsov_F200} as the energy for which
 the scaling ratio $K=F(>$$E^*)/A$ is approximately constant in a wide range of spectral parameters.
Here $F(>$$E)$ is the integral omnidirectional flux of protons with energy above $E$; the activity $A$ of
 the radioisotope (quantified via the disintegration rate per minute per kilogram, dpm/kg) at a given depth $h$ is computed from
 the production rate $Q$ (Equation~\ref{eq:Q}), assuming a flux of particles constant in time and
 explicitly considering the exact composition, exposure age $T$, and erosion rate $r$ for each sample as
\begin{equation}
A(h)=\int_0^T{Q(h')\cdot\exp{(-t/\tau)}\cdot dt},
\label{Eq:AQ}
\end{equation}
where $h'=h+\rho r t$ is the depth at time $t$  with account of erosion, $\rho$ is the rock's density,
 and $\tau$ is the lifetime of the isotope.
We assumed in computations that the production rate $Q$ and erosion rate $r$ are constant in time.
An example is shown in Figure~\ref{f:K_factor} (panels A and B) for the EXP and POW models (see Eqs.~\ref{eq:exp} and
 \ref{eq:pow}, respectively) for the depth $h=$1 g/cm$^2$.
We can see that  there is a value of $E^*$ at which $K$ is roughly constant in a wide parametric range.
As the range of parameters, we considered $R_0$=[30\,--\,200] MV for the EXP model and $\gamma$=[2\,--\,5] for the POW model,
 which roughly covers the range of values for the observed SEP events during the last decades.
We have checked that the exact choice of the parametric range does not affect the results.
The exact values of $E^*$ were found by minimizing the relative variability
 $\Delta K/K$, where $\Delta K\equiv K_\textrm{max}-K_\textrm{min}$ is the full range of
 the $K$-values within the studied interval of spectral parameters.
For the examples shown in Figures~\ref{f:K_factor}A and B, the values of $E^*$ were found to be 39.6 MeV and 39.1 MeV,
 respectively, for the EXP and POW models, with the corresponding $\Delta K/K\leq 2$\% in both cases.

%Some examples of finding the effective energy $E^*$ are shown in Figures~\ref{f:K_factor}C and D.
%The value of $E^*$ is defined as the position of the minimum of the corresponding $\Delta K/K$ curve,
% which is very sharp, suggesting that the effective energy is well defined.
The resulting ``calibration'' curve of $E^*$ as a function of depth $h$ is shown in panel C for
 the two spectral models.
We can see that the two curves nearly coincide in the range of depths from 0.1 to 3 g/cm$^2$,
 giving a clear relation to the effective energy within $\approx 20$\,--\,60 MeV.
We neglected the uncertainties of the definition of $E^*$ in each case, but instead took
 the difference between the EXP and POW models ($\Delta E^*$), which is less than 1 MeV.
The corresponding scaling factor $K$ is shown in panel D for the two models with the $\Delta K$-uncertainties
indicated.
We can see that the EXP and POW models agree with each other within the uncertainties in the depth range
 0.1\,--\,5 g/cm$^2$ ($E^*$=[20\,--\,80] MeV), but diverge at shallower and deeper depths.
Therefore, there is a nearly one-to-one relation between the depth in the sample and energy $E^*$ of the
 SEP particles;  the measured isotope activity at this depth $A(h)$ can then be directly translated
 into the the integral flux $F(>E^*)$ via the coefficient $K$.

Thus, the method is suitable for a robust and model-independent reconstruction of the particle spectrum in
 the energy range between 20 and 80 MeV.
This provides a straightforward way to reconstruct the SEP spectrum directly from the measured depth profile of
 \al activity in the upper layers of lunar samples.

\subsection{Estimation of uncertainties}
\label{s:uncert}
We considered several sources of uncertainties:

\underline{\textit{Measurements.}}
Measurement errors of the isotope activity $A$ are converted into production rate $Q$ using
 the sample's age and erosion rate.
The ensuing errors of $Q$ are called $\sigma_Q$.
These errors are treated as distributed normally around $Q$ with the corresponding standard deviation.
In addition, there is the error of the GCR fitting $\sigma_\textrm{GCR}$ (see Section \ref{s:GCR}),
 which is also considered as normally distributed.

\underline{\textit{Model.}}
The model uncertainty is related to the definition of the effective energy $E^*$ and the conversion
 coefficient $K$ for the given depth in a sample (see Section \ref{Sec:spec}).
These uncertainties are treated as uniformly distributed within the full-range interval.

\underline{\textit{SEP spectral shape.}}
Model computations were made for two bounding spectral shapes: exponential EXP and power-law POW
 (Eqs. \ref{eq:exp} and \ref{eq:pow} in Section \ref{Sec:spec}).
These two models were considered to have equal probabilities.

\underline{\textit{Erosion rate.}}
The erosion rate for each sample remains an unknown parameter, with the range estimated as
 0\,--\,0.5 and 1\,--\,2 mm/Myr for samples 64555 and 74275, respectively (see Table \ref{t:flux}).
These uncertainties enter conversion of the measured isotopes activities $A$ into the production rates.

\begin{figure*}
\begin{center}
\includegraphics[width=1\textwidth]{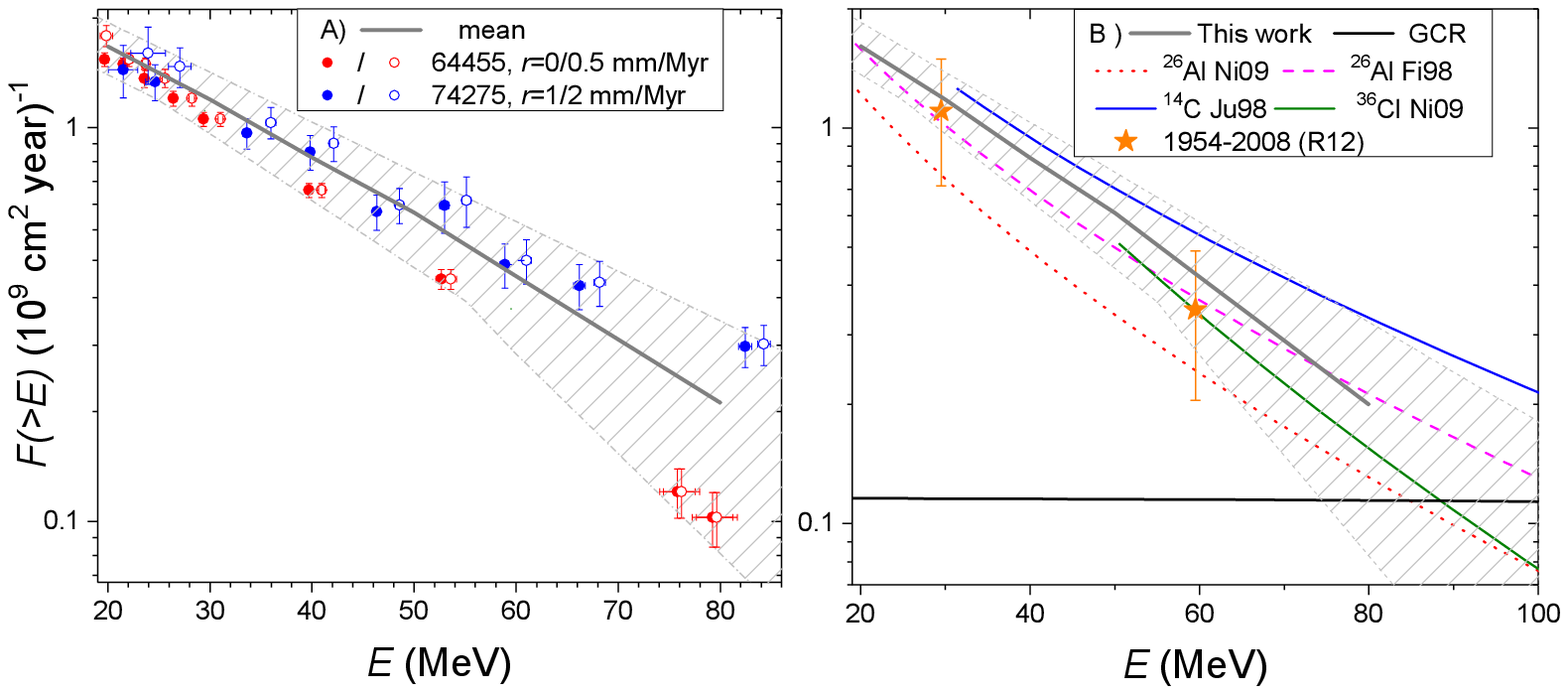}
\end{center}
\caption{
Integral omnidirectional fluxes $F(>$$E)$ of solar energetic protons with energy above the given value $E$.
Panel A: Present reconstruction of the million-year spectrum based on $^{26}$Al in lunar samples.
The red and blue dots depict reconstructions (see Table \ref{t:flux_detail}) from measurements of \al in lunar
 samples 64455 (erosion rate, \textit{r}=0 and 0.5 mm/Myr) and 74275 (erosion rate \textit{r}=1 and 2 mm/Myr),
 respectively, as denoted in the legend.
The thick line and the hatched area depict the average and the full-range uncertainties of the reconstructed fluxes for
 the two samples, considering all the uncertainties, including those of the erosion rate and of the spectral shape.
Exact values are given in Table~\ref{t:flux}.
Panel B: Comparison with other spectra.
The thick  gray line with hatched area is the present reconstruction (identical to that in panel A).
Colored lines represent earlier estimates of the SEP spectra, explicitly assuming the EXP shape (Eq.~\ref{eq:exp}),
 from \al by\citet[][-- Fi98]{fink98}, from $^{14}$C by \citet[][denoted  Ju98]{jull98}, and from \al and $^{36}$Cl
 by \citet[][-- Ni09]{nishiizumi09}.
The orange stars depict the mean values of $F(>$30 MeV) and $F(>$60 MeV) for
 the last solar cycles 1954--2008 \citep{reedy12}, error bars being the standard error of the mean over individual cycles.
The thick black line depicts the GCR contribution ($\phi=496$ MV).
}
\label{f:flux}
\end{figure*}

Uncertainties of the final result were assessed using a Monte Carlo method as described below.
For each data point for a given sample and depth $h$ we made $10^6$ realizations of the reconstructed SEP flux $F(>$$E^*)$.
For each realization we first took a normally distributed random number with the mean $Q(h)$ and standard deviation of
 $\sigma_Q(h)$.
Then the value of the GCR related background was taken in a similar random way using the estimate of
 $\sigma_\textrm{GCR}$.
Thus, the value of $Q_\textrm{SEP}(h)$ was obtained.
Next, we randomly chose the model between exponential and power law, and computed the energy $E^*$
 (from Figure~\ref{f:K_factor}C) taking the scaling factor $K$ as a uniformly distributed
 random number from the interval $K\pm\Delta K$ (from Figure~\ref{f:K_factor}D).
From these obtained $10^6$ values we computed the mean flux $\left< F \right>$ and the uncertainties
 defined as the full range of $E^*$ and the upper and lower 16th percentiles of the $F$ distribution.
The final result is shown in Figure \ref{f:flux} and listed in Table \ref{t:flux}.
The gray-shaded area in Figure \ref{f:flux} denotes the full range of the reconstructed spectrum,
 which is an envelope of individual reconstructions.

\section{Results}

%.........................
\subsection{Galactic cosmic rays on the mega-year timescale}
\label{s:GCR}

Since lunar rocks are bombarded by both SEPs and GCRs, we need first to remove the contribution of GCRs  into the isotope
 production before reconstructing SEPs.
GCRs are much more energetic, but less abundant in the lower energy range than SEPs.
They penetrate much deeper into matter, where initiate nucleonic cascades and dominate the isotope production at
 depths $>20$ g/cm$^2$.

\begin{figure}
\begin{center}
\includegraphics[width=1\columnwidth]{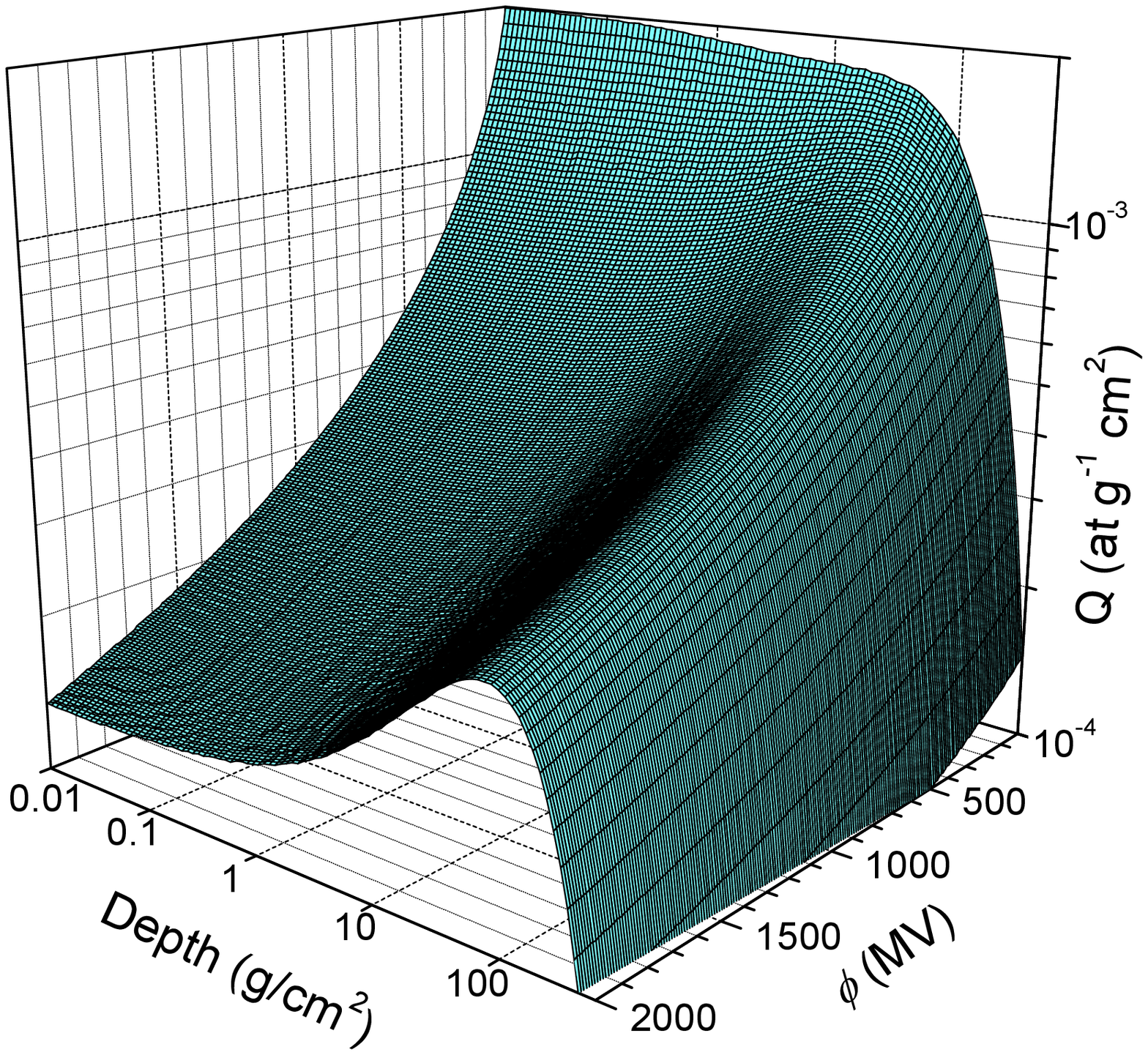}
\end{center}
\caption{
Modeled production $Q$ of cosmogenic \al by galactic cosmic rays in a lunar soil as a function of
 the modulation potential $\phi$ and depth $h$.
The production was computed for the slab geometry and the chemical composition of the Apollo 15 deep drill core \citep{gold77}.
}
\label{f:Q_model}
\end{figure}

\begin{figure*}
\begin{center}
\includegraphics[width=1\textwidth]{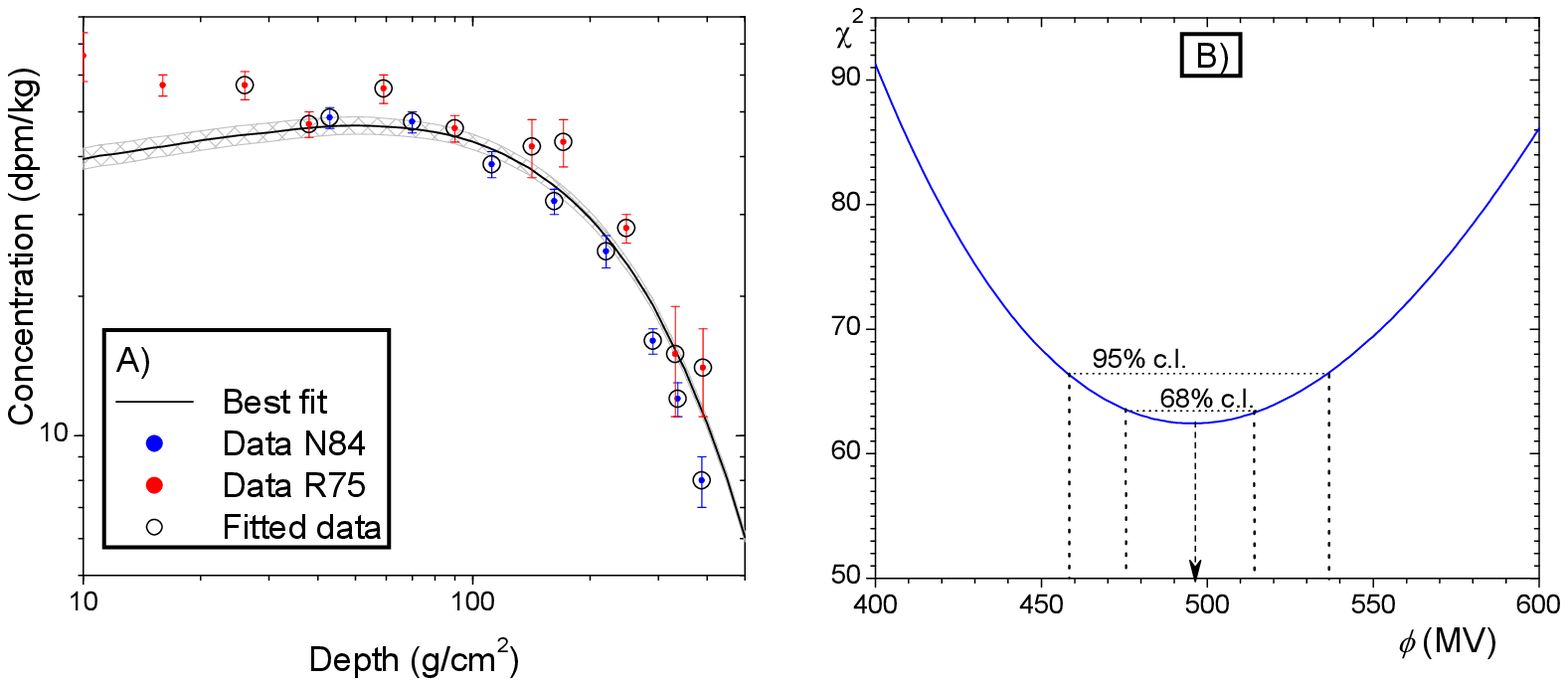}
\end{center}
\caption{
Determination of the mean GCR level.
Panel A) Measured (blue \citep[N84;][]{nishiizumi84} and red \citep[R75;][]{rancitelli75} dots) and the
 best-fit GCR-induced activity of \al in the Apollo 15 deep drill core.
Datapoints used for the fit are  in black circles.
Since activity at depths shallower than 20 g/cm$^2$ can be affected by SEPs, shallow points were not used in the fit.
The gray hatched area denotes the 68\% confidence interval defined in panel B.
Panel B)  $\chi^2$ statistics for the fit of the modeled \al activity shown in Panel A as a function of
the modulation potential $\phi$.
The best-fit $(\chi_{\rm min}^2=62.4)$ value of $\phi=496$ MV is shown by the arrow, while the 68\% (defined as
 $\chi^2_\textrm{min}+1.0$) and 95\% ($\chi^2_\textrm{min}+4.0$) confidence intervals are
 denoted by the dotted lines.
}
\label{f:Q_meas}
\end{figure*}

We adopted a general description of the GCR spectral shape in the form of a force-field approximation
 \citep{gleeson68,caballero04}, which is a common and well validated way to describe
 the long-term variability of GCRs \citep{usoskin_Phi_05,herbst10}.
The near-Earth energy spectrum $J_i(E)$, of GCR particles of type $i$ (proton or heavier species), characterized by
 the charge $Z_i$ and mass $A_i$ numbers, can be presented via the unmodulated local interstellar
 spectrum (LIS) $J_{\textrm{LIS},i}$ and the so-called modulation potential $\phi$ as
\begin{equation}
J_i(E,\phi) = J_{\textrm{LIS},i}(E+\Phi_i)
    \frac{E(E+2E_r)}
    {(E+\Phi_i)(E+\Phi_i+2E_r)},
\end{equation}
where $E$ is the particle's kinetic energy per nucleon, $\Phi_i = \phi(eZ_i/A_i),$ and $E_r=0.938$ GeV/nucleon is the rest mass of a proton.
The energy spectrum of GCR is defined by a single parameter, the modulation potential $\phi$, which quantifies modulation of
 GCR inside the heliosphere by solar magnetic activity.
The reference local interstellar spectrum (LIS) of protons was taken in the form provided by \citet{vos15}
 considering direct in situ measurements of low-energy GCR beyond the termination
 shock and near-Earth measurements in high-energy range.
Results, based on other LIS models, can be straightforwardly converted \citep{asvestari_JGR_17}.
The nucleonic fraction of helium and heavier nuclei was standardly considered as 0.3 to protons in LIS \citep[e.g.,][]{usoskin_Bazi_11}.
We assumed that LIS is constant on a million-year timescale.
Figure~\ref{f:Q_model} shows dependence of the \al production by GCR in the deep drill core as a function of the depth $h$ and
 the modulation potential $\phi$.
Applying this dependence, we fitted the measured concentrations of \al in the Apollo 15 deep core \citep{nishiizumi84,rancitelli75} to
 find the best-fit values of the modulation potential (Figure \ref{f:Q_meas}).
The fit was performed by minimizing $\chi^2$ as shown in panel B.
The best-fit value of the modulation potential was found to be $\phi=496\pm40$ MV for the last million years,
 which is considered hereafter.

This value is significantly smaller than the mean modulation potential $\phi=660\pm20$ MV for the period 1951\,--\,2016
 \citep{usoskin_gil_17}, but consistent within uncertainties with the mean modulation over the Holocene \citep{usoskin_AA_16}
 $\phi=449\pm 70$ MV (reduced to the same LIS).
This highlights the fact that the second half of the 20th century was characterized by MGM with unusually high solar activity.

\subsection{Mean SEP spectrum estimate}

The corresponding GCR contribution was subtracted from the measured isotope content, and the remaining activity
 was ascribed to SEP, whose spectrum was further reconstructed using the ``calibration'' curves shown in
 Figure~\ref{f:K_factor}C and D.
The final reconstruction of the SEP spectrum is shown in Figure~\ref{f:flux}A for the two lunar samples
 and for a range of erosion rates.
Uncertainties were calculated as described in \ref{s:uncert}.
The full range uncertainty (the envelope over different samples and erosion rates) is shown by the gray hatched area.
The reconstructed fluxes lie very close to each other in the energy range of 20\,--\,35 MeV, being almost independent
 on the exact sample, spectral shape, and erosion rate, with the full-range relative uncertainty
 (including the model uncertainty) being within 20\%.
This energy range corresponds to the $F(>$30 MeV), which is often considered as the radiation environment.
Thus, the $F(>$30 MeV) flux is quite robustly defined here.
The uncertainty grows with energy, reaching 40\% at 60 MeV and a factor of two at 80 MeV.
The summary spectrum, which includes the mean and the full-range uncertainties (gray area in Figure~\ref{f:flux}),
 is given in Table~\ref{t:flux} and serves as our final estimate of the mean SEP spectrum over million of years.

\begin{figure*}
\begin{center}
\includegraphics[width=1\textwidth]{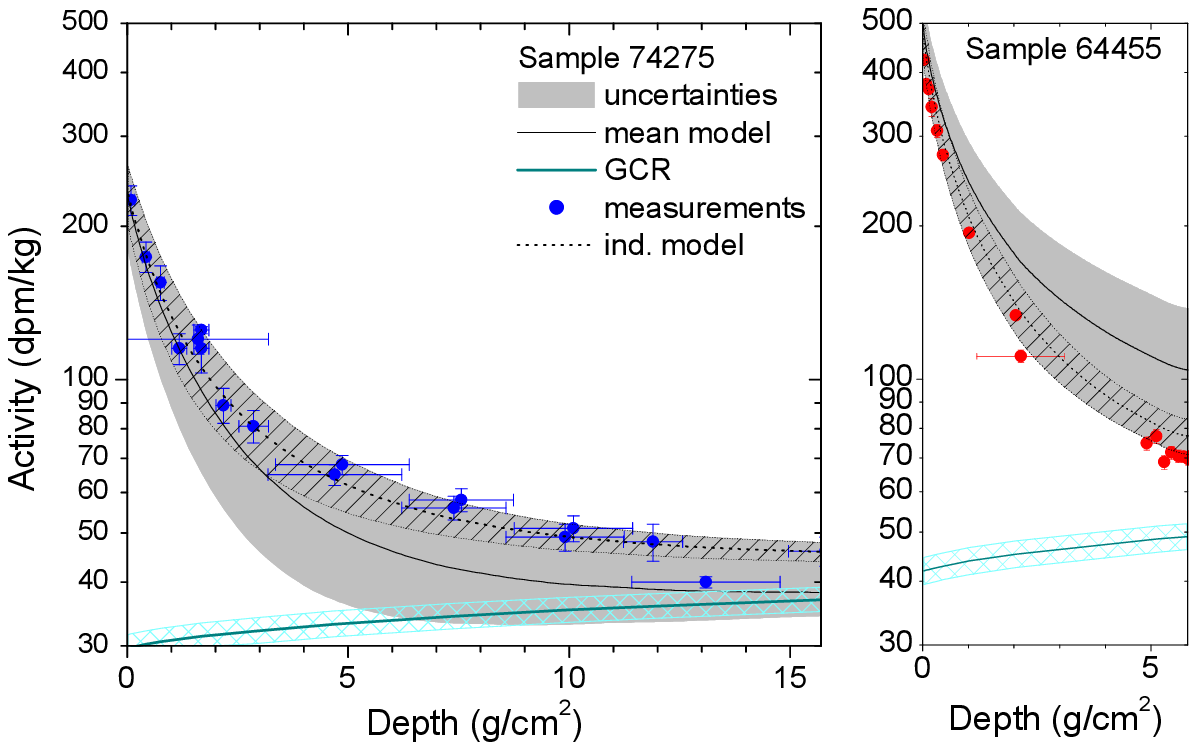}
\end{center}
\caption{
Comparison of the measured \al activity in the analyzed samples 74274 \citep[blue points in panel A, from][]{fink98}
 and 64455 \citep[red points in panel B, from ][]{nishiizumi09} with the values computed here using the final reconstructed
 spectrum (Table~\ref{t:flux} in the main text) as shown by the solid black line (the mean activity) with
 the full-range uncertainty (gray area).
The dashed curves with the hatched areas correspond to spectra reconstructed for individual samples (Table~\ref{t:flux_detail}).
The light blue hatched area with the olive curve depicts the GCR contribution to the activity.
}
\label{f:Q_recon}
\end{figure*}

As a consistency check, we performed the following computation:
 from the spectra reconstructed above (both GCR and SEP) we calculated, using realistic geometry and
 chemical compositions, the expected activity of \al in a lunar rock and compared it with the actual measurements
 (Figure~\ref{f:Q_recon}).
The computed activity appears consistent with the data within the uncertainties, and lies between the two samples.
Although it does not prove the correctness of the method, it does imply that the model is free from internal systematic errors.

A comparison of the new reconstructed million-year SEP spectrum with earlier estimates and the direct space-era data
 is shown in Figure~\ref{f:flux}B.
Color curves present earlier estimates of the SEP spectrum \citep{fink98,jull98,nishiizumi09} explicitly
 assuming its EXP shape (Equation~\ref{eq:exp}).
They all agree within a factor of two and fit  into the uncertainty of the
 present reconstruction (except for the \al-based Ni09).
The curves are shown in the energy range corresponding to the production threshold (see Figure~\ref{f:prod_eff}B),
 since their extension to the lower energy range would be based on an extrapolation of the exponential shape
 but not on data.
The SEP spectrum is more uncertain for energies above 100 MeV, where the contribution of GCR becomes dominant.
We conclude that the present direct reconstruction of the mean SEP spectrum is consistent with earlier
 estimates which were obtained using an explicit {a priori} assumption on the EXP shape.

For comparison, we also show  the mean integral fluxes $F(>$30 MeV) and $F(>$60 MeV) for the last solar cycles 1954\,--\,2008
 \citep{reedy12} as orange stars.
The flux reconstructed for the last million years agrees, within the uncertainties, with that for the last cycles.
This is an interesting result since solar activity (the modulation potential $\phi$)
 was significantly higher during the recent decades due to the Modern grand maximum than in the past (Section~\ref{s:GCR}).
This implies that the mean SEP flux does not depict a notable dependence on the overall solar activity level.
This result is concordant with the fact that strongest known extreme SEP events \citep{miyake12,miyake13,usoskin_775_13,jull14,wang17},
 such as the events of 775AD, 993AD, BC3372, and the Carrington event of 1859, occurred during periods of moderate solar activity.
This provides  new observational constraints \citep{hudson10}.
\begin{table}
\caption{
Reconstructed integral omnidirectional flux $F(>$$E^*)$ in $10^9$ (cm$^2$ year)$^{-1}$ based on measurements of \al in
 the two samples (see Section~\ref{Sec:data}) and boundary erosion rates $r$.
These data are shown in Figure \ref{f:flux} of the main text.
}
\begin{center}
\begin{tabular}{|c|c|c|c|}
\hline
\multicolumn{4}{|c|}{Sample 64455}\\
\multicolumn{2}{|c|}{$r=0$ mm/Myr} & \multicolumn{2}{|c|}{$r=0.5$ mm/Myr} \\
$E^*$(MeV) & $F(>E^*)$ & $E^*$(MeV) & $F(>E^*)$ \\
\hline
$16.8\pm0.8 $  &  $ 1.71\pm0.14 $  &  $ 19.9\pm0.6 $  &  $ 1.71\pm0.11 $  \\
$19.7\pm0.4 $  &  $ 1.47\pm0.06 $  &  $ 22.1\pm0.4 $  &  $ 1.49\pm0.06 $  \\
$21.5\pm0.3 $  &  $ 1.43\pm0.05 $  &  $ 23.7\pm0.3 $  &  $ 1.45\pm0.05 $  \\
$23.6\pm0.3 $  &  $ 1.32\pm0.07 $  &  $ 25.6\pm0.3 $  &  $ 1.34\pm0.07 $  \\
$26.4\pm0.2 $  &  $ 1.18\pm0.05 $  &  $ 28.2\pm0.3 $  &  $ 1.19\pm0.05 $  \\
$29.4\pm0.2 $  &  $ 1.05\pm0.04 $  &  $ 31.0\pm0.2 $  &  $ 1.05\pm0.04 $  \\
$39.7\pm0.3 $  &  $ 0.70\pm0.03 $  &  $ 41.0\pm0.3 $  &  $ 0.69\pm0.03 $  \\
$52.7\pm0.6 $  &  $ 0.42\pm0.02 $  &  $ 53.6\pm0.6 $  &  $ 0.41\pm0.02 $  \\
$75.8\pm1.8 $  &  $ 0.12\pm0.02 $  &  $ 76.2\pm1.8 $  &  $ 0.12\pm0.02 $  \\
$79.2\pm2.0 $  &  $ 0.10\pm0.02 $  &  $ 79.7\pm2.0 $  &  $ 0.10\pm0.02 $  \\
\hline
\multicolumn{4}{|c|}{Sample 74275}\\
\multicolumn{2}{|c|}{$r=1$ mm/Myr} & \multicolumn{2}{|c|}{$r=2$ mm/Myr} \\
$E^*$(MeV) & $F(>E^*)$ & $E^*$(MeV) & $F(>E^*)$ \\
\hline
 $ 21.5\pm1.5 $  &  $ 1.41\pm0.21 $  &  $ 23.9\pm1.7 $  &  $ 1.55\pm0.26$  \\
 $ 24.6\pm0.9 $  &  $ 1.31\pm0.14 $  &  $ 27.0\pm1.1 $  &  $ 1.43\pm0.17$  \\
 $ 33.6\pm0.4 $  &  $ 0.97\pm0.09 $  &  $ 36.0\pm0.5 $  &  $ 1.03\pm0.09$  \\
 $ 39.8\pm0.4 $  &  $ 0.87\pm0.09 $  &  $ 42.1\pm0.4 $  &  $ 0.91\pm0.09$  \\
 $ 46.3\pm0.4 $  &  $ 0.61\pm0.06 $  &  $ 48.6\pm0.4 $  &  $ 0.64\pm0.06$  \\
 $ 53.0\pm0.5 $  &  $ 0.63\pm0.09 $  &  $ 55.1\pm0.5 $  &  $ 0.65\pm0.10$  \\
 $ 58.9\pm0.5 $  &  $ 0.45\pm0.06 $  &  $ 61.0\pm0.5 $  &  $ 0.46\pm0.06$  \\
 $ 66.2\pm0.6 $  &  $ 0.40\pm0.05 $  &  $ 68.2\pm0.6 $  &  $ 0.41\pm0.05$  \\
 $ 82.4\pm0.6 $  &  $ 0.28\pm0.03 $  &  $ 84.3\pm0.6 $  &  $ 0.28\pm0.03$  \\
\hline
\end{tabular}
\end{center}
\label{t:flux_detail}
\end{table}

\begin{table}
\caption{Reconstructed integral omnidirectional flux of SEP over the last million years (see also Figure~\ref{f:flux}):
 the formal mean flux and its full-range uncertainty, both in units of $10^9$ (cm$^2$ year)$^{-1}$.
}
\begin{center}
\begin{tabular}{|c|c|c|}
\hline
$E$ (MeV) & $\left<F(>E)\right>$ & $\Delta F$ \\
\hline
20      & 1.61  & 0.24 \\
30      & 1.18  & 0.21 \\
40      & 0.84  & 0.19 \\
50      & 0.61  & 0.17 \\
60      & 0.42  & 0.16 \\
70      & 0.29  & 0.14 \\
80      & 0.20  & 0.12 \\
\hline
\end{tabular}
\end{center}
\label{t:flux}
\end{table}

\subsection{Estimate of the SEP event occurrence probability}

Following the idea presented by \citet{kovaltsov14}, we can assess the occurrence probability density function (OPDF)
 of SEP events, based on the reconstructed spectrum.
Since only the mean flux of SEP can be directly estimated from lunar data, further modeling is needed to assess
 the occurrence rate of individual events.
For example, the entire mean flux can be produced by a huge single event that occurred a while ago.
Such an extreme assumption was made, for example,  by \citet{reedy96}, but it is obviously unrealistic since it assumes
 that there are no other weaker events during the lifetime of the isotope.
In reality, there is always a distribution of events over their strength and occurrence rate.
A more appropriate estimate can be made by assuming such a realistic distribution.
\begin{figure}
\centering
\includegraphics[width=\columnwidth]{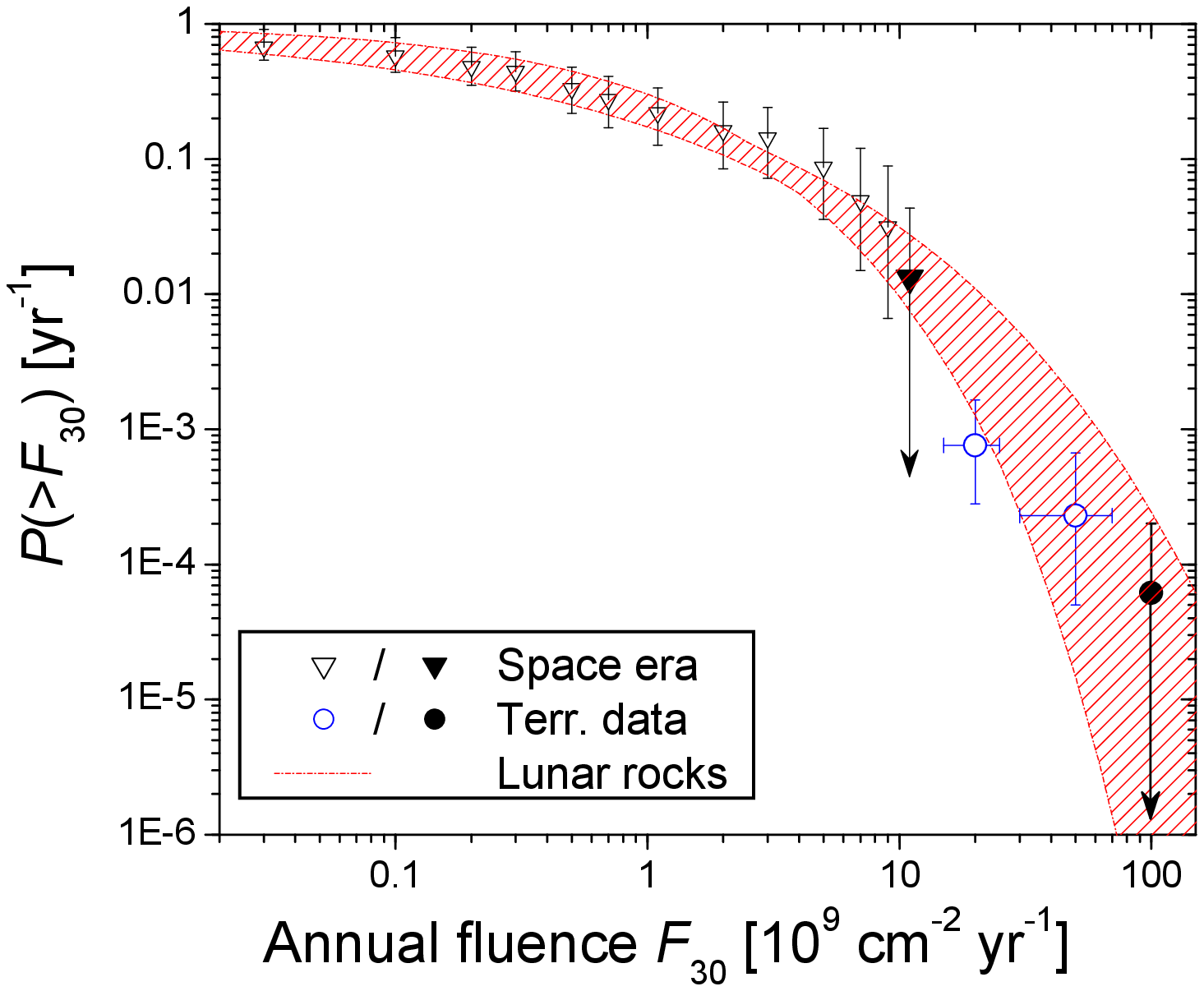}
\caption{Occurrence probability density function (OPDF) of solar energetic particles with the annual $F(>$30 MeV) fluence exceeding the
 giving value.
 The triangles denote OPDF based on the data for the space era \citep{usoskin_ApJ_12}.
 The circles correspond to the SEP events estimated from the terrestrial cosmogenic isotope data \citep[modified after][]{usoskin_LR_17}.
 Open symbols indicate the measured/estimated values, filled symbols indicate  a conservative upper bound.
 Error bars bound the 68\% confidence interval.
 The red hatched area encompasses the OPDF estimated in this work from \al in lunar samples.
 }
\label{Fig:IPDF}
\end{figure}
Let us assume that OPDF of SEP annual fluence can be approximated by the
 Weibull distribution \citep{weibull51}, which is often used to describe the occurrence probability of solar events
 \citep[see, e.g.,][and discussion therein]{gopalswamy18}
\begin{equation}
P(>F) = \exp{\left(-\, ({F/F_0})^k\right)},
\label{Eq:wei}
\end{equation}
where $P$ is the probability that  a SEP event with the fluence greater than $F$ will occur within one year, and
 $k$ and $F_0$ are two fitted parameters of the model.
Then the mean fluence $\langle F\rangle$ is defined as
\begin{equation}
\langle F \rangle \equiv \int{F\,{dP\over dF}\,dF}.
\label{Eq:F}
\end{equation}
Now this model can be fitted to data using several observational constraints:
\begin{itemize}
\item
The mean $F(>$30 MeV) flux (Equation~\ref{Eq:F}) takes the values between 1 and 1.4  $\cdot 10^{9}$ (cm$^2$ yr)$^{-1}$
 as corresponding to the mean reconstructed values (Table~\ref{t:flux});

\item
The fact that no events with $F(>$30 MeV) exceeding $10^{11}$ (cm$^2$ yr)$^{-1}$ have been found and are unlikely to
 be found over the Holocene \citep{usoskin_LR_17} poses an upper limit of the corresponding OPDF as
 $P(>$$10^{11}\,{\rm cm}^{-2})\leq 2.1\cdot 10^{-4}$ yr$^{-1}$;

\item
A lower limit of $P(>$2$\cdot 10^{10}\,{\rm cm}^{-2})\geq 2.8\cdot 10^{-4}$ yr$^{-1}$ is set that  corresponds
 to the three historical events of 3372 BC \citep{wang17}, 775 AD \citep{miyake12}, and 994 AD \citep{miyake13} securely found
  during the Holocene, but there may be more similar events discovered \citep[e.g.,][]{park17} since  the entire record
  has not been analyzed yet \citep{usoskin_LR_17}.
\end{itemize}
The suitable parameter range was found as $F_0=$[0.2\,--\,0.8] $10^9$ cm$^{-2}$yr$^{-1}$ and $k=$[0.3\,--\,0.57]
 with a high correlation between them.
The corresponding distribution function is shown as the red hatched area in Figure~\ref{Fig:IPDF} and describes well
 all the data points including the data from lunar rocks.
It is interesting to note that the first blue dot in Figure~\ref{Fig:IPDF} lies too low versus the reconstructed OPDF,
 implying that the number of known events with $F(>30$ MeV)$\geq 2\cdot 10^{10}$ (cm$^2$ yr)$^{-1}$ is smaller than expected
 and that we expect more events of this strength to be discovered over the Holocene \citep{park17}.
%Overall, a very sharp roll-off of the distribution is observed, suggesting that probably, during the event of 775 AD,
% Sun has approached close to its limit of producing extreme SEP events.
According to this estimate, no events with $F(>$30 MeV)$>5\cdot 10^{10}$ and $\approx 10^{11}$ protons/cm$^2$ are expected
 on  timescales of a thousand and a million years, respectively.

\section{Discussion and conclusions}

We have shown that activity of \al measured in the upper shallow layers ($<$5 g/cm$^2$) of lunar rocks serves
 as a good integral particle spectrometer in the energy range of 20\,--\,80 MeV.
The lower bound is limited by the threshold of the isotope's production reactions.
On the other hand, more energetic particles with energy above 100 MeV can initiate nucleonic cascades in deeper layers
 making the yield function of the isotope's production  increase with energy (Figure~\ref{f:yf}B), which distorts the
 characteristics of an integral spectrometer.
Moreover, the GCR contribution to the isotope production grows with depth (Figure~\ref{f:flux}), which makes
 separation of the SEP signal less robust in deeper layers.
Thus, only the SEP flux in the energy range 20\,--\,80 MeV can be reliably reconstructed.

Although the idea of reconstruction of the spectrum of solar energetic particles from lunar samples has been
 exploited in the past, a new approach based on precise computations of the yield function makes it possible to
 use lunar rocks  as a good particle spectrometer able to estimate the SEP spectrum without any {a priori}
 assumptions on its shape.
Using this method, we  provided the first realistic model-independent reconstruction of the SEP energy
 spectrum in the energy range 20\,--\,80 MeV on a timescale of a million years and showed that it is
 consistent with earlier estimates and with the modern values.
We also evaluated, using the \al data from deep layers, the million-year mean flux of GCR, which appears to be
 significantly less modulated (the modulation potential $\phi=496\pm 40$ MV) than during the last decades
 ($\phi=660\pm 20$ MV), confirming the importance of MGM of solar activity in the second
 half of the twentieth century.
The fact that the mean million-year SEP spectrum is consistent with the modern values, obtained during the MGM,
 implies a lack of notable dependence of the SEP flux on the level of solar activity,
 consistent with the fact that the strongest known historical solar event occurred during periods of moderate
 solar activity.
This puts new observational constraints on solar physics and becomes crucially important for
 assessing radiation hazards for the planned space missions.

%.........................
\begin{acknowledgements}
This work was supported by the Center of Excellence ReSoLVE (Project 272157) of the Academy of Finland.
The authors are thankful to Anton Artamonov for the help in verification of the Geant4 model of lunar samples.
\end{acknowledgements}

%.........................

%\section*{References}

%\bibliographystyle{elsarticle-harv}
%\bibliography{bibliography.bib}
%\bibliography{J:/usoskin/papers/usoskin_all}
%\bibliography{C:/DATA__/papers/usoskin_all}

\end{document}